\documentclass{article}
\usepackage{frascatiphys}
\usepackage{graphicx}
\usepackage{amsmath, amssymb, latexsym}
\def\vev#1{\left\langle #1\right\rangle}

\begin{document}
\title{Discrete Dark Matter Model And Reactor Mixing Angle}
\author{
J. M. Lamprea\footnote{jmlamprea@fisica.unam.mx}\\
{\em  Institute of Physics,  UNAM, A.P. 20-364, Mexico City 01000, Mexico}}
\maketitle{}
\baselineskip=11.6pt
\begin{abstract}
We present a scenario where the stability of dark matter and the phenomenology of neutrinos are related by the breaking of a flavour symmetry. We propose two models based on this idea for which we have obtained interesting neutrino and dark matter phenomenology. 
\end{abstract}
\baselineskip=14pt
\section{Introduction}
We propose an extension of the SM\cite{Lamprea:2016egz} in the context of the discrete dark matter (DDM) mechanism\cite{Hirsch:2010ru}.  This mechanism is based upon the fact that the breaking of a discrete non-Abelian flavour symmetry accounts for the neutrino masses and mixing pattern and for the dark matter stability. In the original DDM model $A_4$ is considered as the flavour symmetry and the particle content includes four scalar $SU(2)$ doublets: three in the $A_4$ triplet $\eta = (\eta_1,\, \eta_2,\, \eta_3)$ and the SM Higgs $H$ as a singlet, four right-handed neutrinos, three of them in a triplet $N_T = (N_1, N_2, N_3)$ and $N_4$ as a singlet. On the other hand, the charged leptons (doublets $L_i$ and singlets $l_i$) are assigned to the three different singlets of $A_4$ in such a way that their mass matrix is diagonal. Finally in the DDM model, the breaking of $A_4$ into $Z_2$, through the electroweak symmetry breaking, provides the stability mechanism for the DM and accounts for the neutrino masses and mixing patterns by means of the type I seesaw. Nevertheless the original DDM model predicts an inverse mass hierarchy, a massless neutrino and a vanishing reactor neutrino mixing angle that nowadays ruled out\cite{Tang:2015vug}. 
\section{Reactor mixing angle and the DDM mechanism}
\label{sec:Models}
We consider two extensions (model A and B) of the original DDM model, where we have added  one extra RH neutrino $N_5$, as $\bf{1^{\prime}}$ in model A and $\bf{1^{\prime\prime}}$ in model B, and three real scalar singlets of the SM as the triplet $\phi= (\phi_1,~\phi_2,~ \phi_3) $.  The relevant particle content is summarised on Tables  \ref{tab:ModA} and \ref{tab:ModB}.  The flavon fields $\phi $ acquire a vev around the seesaw scale, such that  $A_4$ is broken into $Z_2$ at this scale  contributing to the RH neutrino masses.
\subsection*{Model A}
\begin{table}[h!]
\begin{center}
\begin{tabular}{|c|c|c|c|c|c|c|c|c|c||c|c|c|}
\hline
& $\,L_e\,$ & $\,L_\mu \,$ & $\,L_\tau \,$ & $\,\,l_e^c\,\,$ & $\,\,l_\mu^c\,\,$ & $\,\,l_\tau^c\,\,$ & $N_T\,$ & $\,N_4\,$ & $\,N_5\,$ & $\,H\,$ & $\,\eta \,$ & $\,\phi \,$\\
\hline
SU(2) & 2 & 2 & 2 & 1 & 1 & 1 & 1 & 1 & 1 & 2 & 2 & 1\\
\hline
$A_4$ & 1 & $1^\prime$ & $1^{\prime\prime}$ & 1 & $1^{\prime\prime}$ & $1^\prime$ & 3 & 1 & $1^\prime$ & 1 & 3  &  3 \\
\hline
\end{tabular}\caption{\it Summary of the relevant particle content for model A.} \label{tab:ModA}
\end{center}
\end{table}
Considering the matter content in Tab. \ref{tab:ModA},  the relevant part of the Lagrangian is given by\footnote{The term $y_1^N [N_T \,\phi]_3 N_T$ accounts for the symmetric part of $[N_T \,\phi]_{3_1}$ and $[N_T \,\phi]_{3_2}$.}
\footnote{ $[a,\,b]_{j}$ stands for the product of the two triplets $a, \,b$  are contracted into the $j$  representation of $A_4$}:
\begin{align} 
\label{eq:YukA}
\mathcal{L}^{(\text{A})}_{\text{Y}} &= 
y_1^\nu L_e [N_T\, \eta]_1 + y_2^\nu L_\mu [N_T\, \eta] _{1''} + y_3^\nu L_\tau [N_T\, \eta]_{1'} + y_4^\nu L_e \,N_4\, H\nonumber\\ 
&+ y_5^\nu L_\tau\, N_5\, H\ + M_1\, N_T N_T + M_2\, N_4 N_4 \\
&+ y_1^N [N_T \,\phi]_3 N_T +  y_2^N [N_T\, \phi]_{1} N_4 + y_3^N [N_T\, \phi]_{1''} N_5 + \textit{h.c.}\nonumber
\end{align}
In this way $H$ is responsible for the quarks and charged lepton masses, the latter automatically diagonal. The Dirac neutrino mass matrix arises from $H$ and $\eta$, and the flavon fields contribute to the RH neutrino mass matrix. In order to preserve a $Z_2$ symmetry, the alignment of the vev's take the form:
\begin{equation}
\label{eq:Vevs}
\vev{H^0} = v_h \ne 0,~
\vev{ \eta^0_1} = v_\eta \ne 0,~
\vev{\eta^0_{2,3}} = 0,  ~
\vev{ \phi_1} = v_\phi \ne 0,~
\vev{\phi_{2,3}} = 0.
\end{equation} 

From Eqs. (\ref{eq:YukA}) and~(\ref{eq:Vevs}) the light neutrinos get Majorana masses through the type I seesaw relation taking the form:
\begin{equation} 
\label{eq:MnuA}
m_\nu^{(\text{A})}
\equiv
\begin{pmatrix}
a & 0 & b\\
0 & 0 & c\\
b & c & d
\end{pmatrix},
\end{equation}
with $a = \frac{(y_4^\nu v_h)^2 }{M_2}$, $b = \frac{y_1^\nu y_5^\nu v_\eta v_h }{y_3^N v_\phi} - \frac{ y_2^N y_4^\nu y_5^\nu v_h^2 }{y_3^N M_2 }$,  $c = \frac{y_2^\nu y_5^\nu v_\eta v_h}{y_3^N v_\phi}$, and
$ d = \frac{ (y_2^N y_5^\nu v_h)^2 }{(y_3^N)^2 M_2 } - \frac{ (y_5^\nu v_h)^2  M_1}{(y_3^N v_\phi)^2} + 2\ \frac{y_3^\nu y_5^\nu v_\eta v_h}{y_3^N v_\phi}$. The mass matrix in Eq. (\ref{eq:MnuA}) has the $B_3$ two-zero texture\cite{Frampton:2002yf} which is consistent with both neutrino mass hierarchies and can accommodate the experimental value for the reactor mixing angle, $\theta_{13}$\cite{Meloni:2014yea}.
\subsection*{Model B}
\begin{table}[h!]
\begin{center}
\begin{tabular}{|c|c|c|c|c|c|c|c|c|c||c|c|c|}
\hline
& $\,L_e\,$&$\,L_{\mu}\,$&$\,L_{\tau}\,$&$\,\,l_{e}^c\,\,$&$\,\,l_{{\mu}}^c\,\,$&$\,\,l_{{\tau}}^c\,\,$&$N_{T}\,$&$\,N_4\,$&$\,N_5\,$&$\,H\,$&$\,\eta\,$&\,$\phi\,$\\
\hline
SU(2) & 2 & 2 & 2 & 1 & 1 & 1 & 1 & 1 & 1 & 2 & 2 & 1\\
\hline
$A_4$ & 1 & $1^\prime$ & $1^{\prime\prime}$ & 1 & $1^{\prime\prime}$ & $1^\prime$ & 3 & 1 & $1^{\prime\prime}$ & 1 & 3 & 3\\
\hline
\end{tabular}
\caption{\it Summary of the relevant particle content for model B.}\label{tab:ModB}
\end{center}
\end{table}
The relevant part of the Lagrangian for model B, Tab.~\ref{tab:ModB}, is given by
\begin{align}
\label{eq:YukB}
\mathcal{L}^{(\text{B})}_{\text{Y}} &= 
y_1^\nu L_e [N_T\, \eta]_1 + y_2^\nu L_\mu [N_T\, \eta] _{1''} + y_3^\nu L_\tau [N_T\, \eta]_{1'} + y_4^\nu L_e \,N_4\, H \nonumber\\
&+ y_5^\nu L_\mu\, N_5\, H + M_1\, N_T N_T + M_2\, N_4 N_4 \\
 &+ y_1^N [N_T \,\phi]_{3} N_T + y_2^N [N_T\, \phi]_{1} N_4 + y_3^N [N_T\, \phi]_{1'} N_5 + \textit{h.c.}\nonumber
\end{align} 
The mass matrix of the charged leptons is diagonal, while the light neutrinos Majorana mass matrix after the type I seesaw is
\begin{equation}
 \label{eq:MnuB}
m_\nu^{(\text{B})}
\equiv
\begin{pmatrix}
a & b & 0\\
b & d & c\\
0 & c & 0
\end{pmatrix},
\end{equation}
with $a$ and $b$ as in model A and $c = \frac{y_3^\nu y_5^\nu v_\eta v_h}{y_3^N v_\phi}$, and
$d = \frac{ (y_2^N y_5^\nu v_h)^2 }{(y_3^N)^2 M_2 } - \frac{ (y_5^\nu v_h)^2  M_1}{(y_3^N v_\phi)^2} + 2\ \frac{ y_2^\nu y_5^\nu v_\eta v_h}{ y_3^N v_\phi }$. The mass matrix in Eq. (\ref{eq:MnuB})  correspond to the two-zero texture mass matrix $B_4$\cite{Frampton:2002yf}, which is also consistent with both neutrino mass hierarchies and can also accommodate the reactor mixing angle.
\section{Results}
\label{sec:ModRes}
We performed a numerical scan using the four independent constraints coming from the two complex zeroes, to correlate two of the neutrino mixing parameters.  We use the neutrino oscillation fit data from different groups, for instance\cite{Capozzi:2016rtj}, as inputs and numerically scanned within their $3\sigma$ regions to determine the allowed parameter space. 

\begin{figure}[!h]
\centering
\includegraphics[scale=0.26]{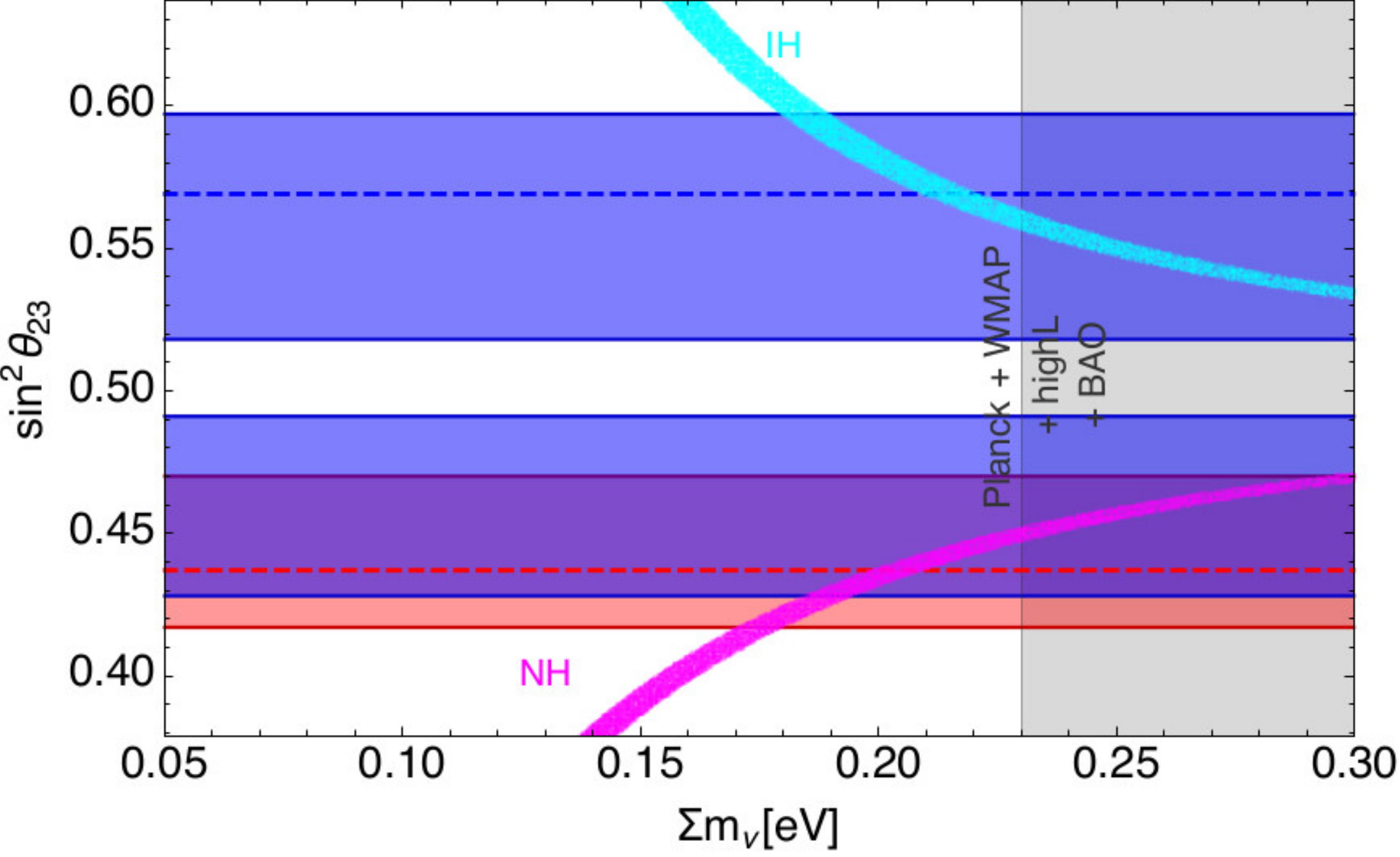} \includegraphics[scale=0.26]{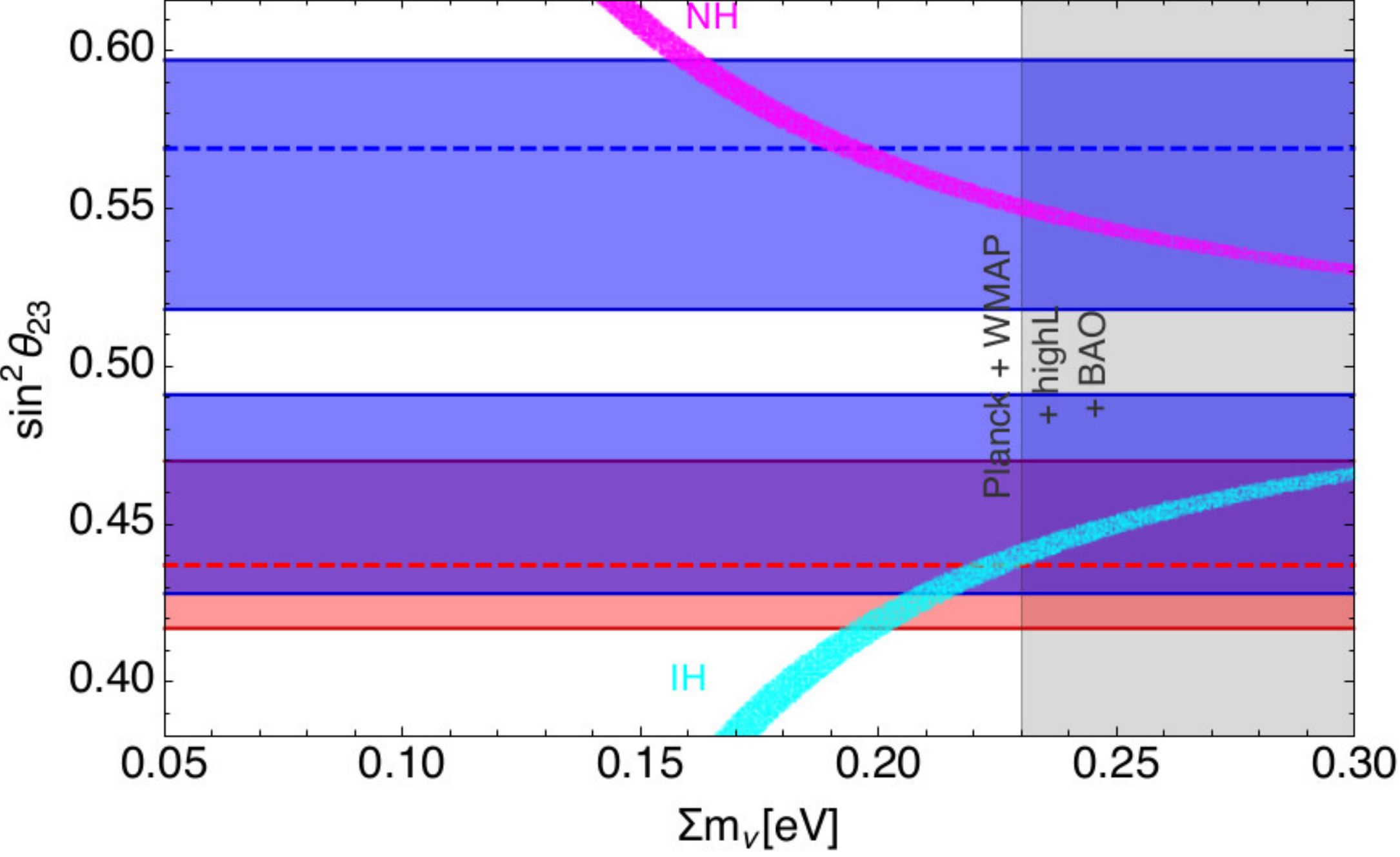}
\caption{Correlation between $\sin^2 \theta_{23}$ and the sum of the light neutrino masses, $\sum m_\nu$, see text for description.}
\label{fig:graf4}
\end{figure}
In Fig.~\ref{fig:graf4} we show the correlation between the atmospheric mixing angle, $\sin^2 \theta_{23}$, and the sum of light neutrino masses, $\textstyle \sum m_\nu$, for model A (B) on the left (right). In the graphics, the allowed $3\sigma$ regions in $\sin^2 \theta_{23}$ vs. $\textstyle \sum m_\nu$,  for the normal hierarchy (NH) is plotted in magenta and for the inverse hierarchy (IH) in cyan. The  $1\sigma$ in the atmospheric angle is represented by the horizontal blue (red) shaded regions for the IH (NH) and the best fit values correspond to the horizontal blue (red) dashed lines for the IH (NH). The grey vertical band represents a disfavoured region in neutrino masses\cite{Ade:2015xua}.  Fig.~\ref{fig:graf4} also shows that in model A both hierarchies have an overlap with the 1$\sigma$ region for $\sin^2 \theta_{23}$, while in model B only in the IH case has such overlap in the second octant.

\begin{figure}[!h]
\centering
\includegraphics[scale=0.26]{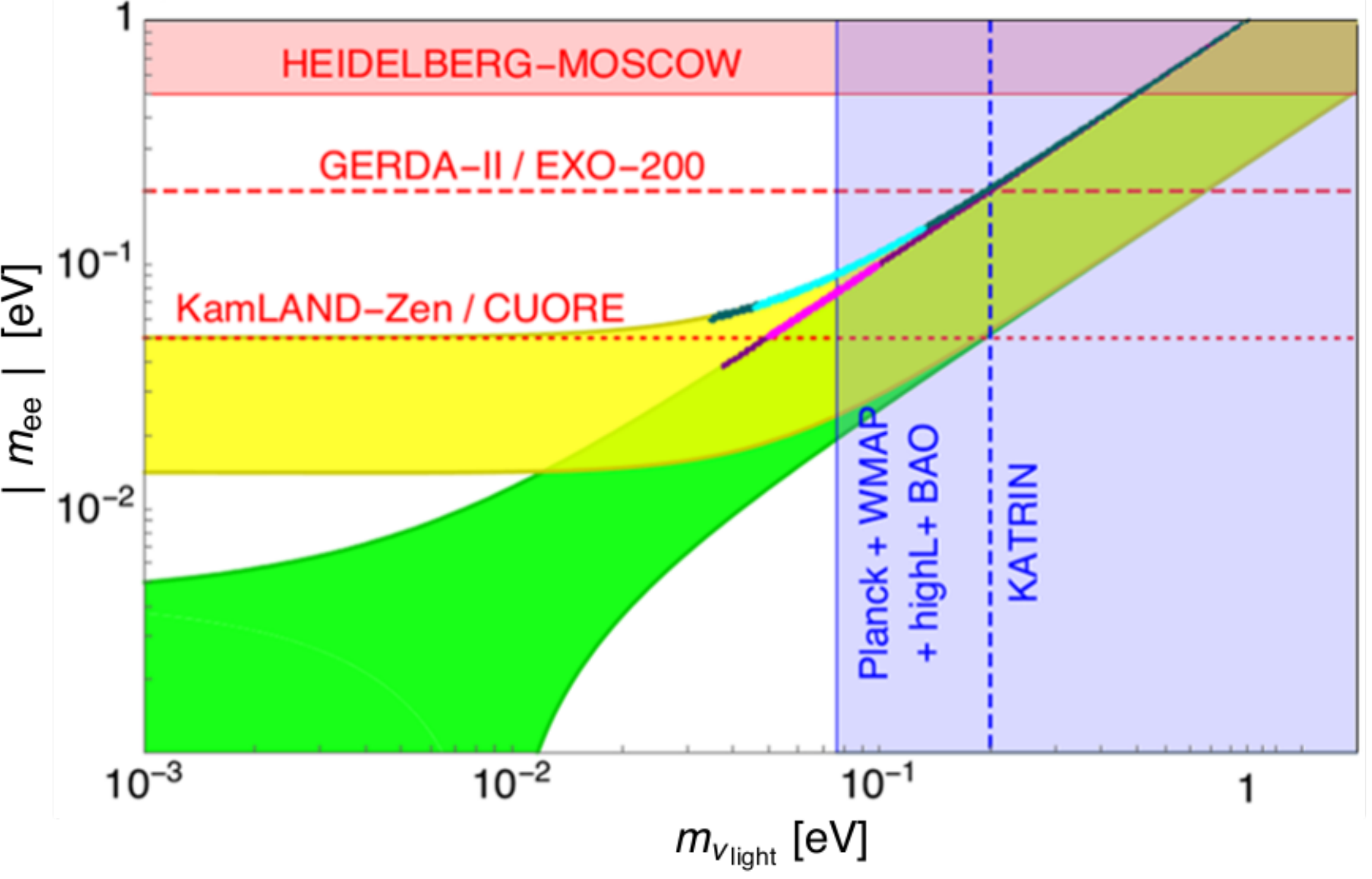} \includegraphics[scale=0.26]{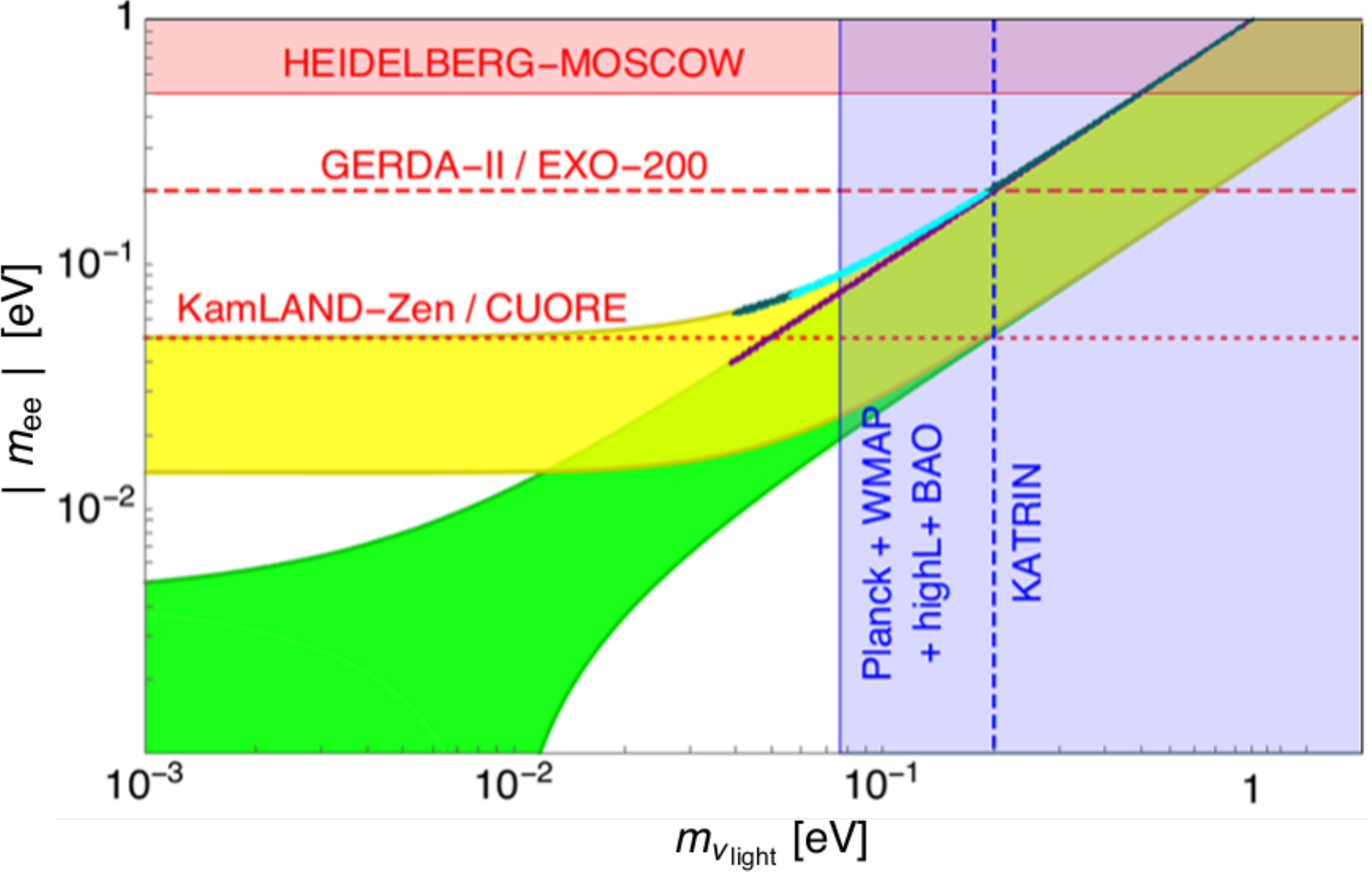}
\caption{Effective $0\nu\beta\beta$ parameter $|m_{ee}|$ versus the lightest neutrino mass $m_{\nu_\text{light}}$, see text for description.}
\label{fig:graf5}
\end{figure}
The Fig.~\ref{fig:graf5} shows $m_{\nu_\text{light}}$ vs. $|m_{ee}|$ for model A (B) on the left (right). The region for the NH (IH) within $3\sigma$ in $\sin^2 \theta_{23}$ are in dark magenta (dark cyan) and the overlap for 1$\sigma$ in magenta (cyan). The red (blue) shaded region corresponds the current experimental limits \cite{Ade:2015xua,neutrinoless}. The yellow (green) the bands correspond to the $3\sigma$ ``flavor-generic" IH (NH) spectra.  The Fig.~\ref{fig:graf5} also shows that the results in model B do not overlap with the 1$\sigma$ region for NH case.  The models predict Majorana phases giving a minimal cancellation for the $|m_{ee}|$.  Both two-zero textures are sensitive to the value of the atmospheric mixing angle that is translated as the localised region for the neutrinoless double beta decay effective parameter within the near future experimental sensitivity.
\section{Conclusions}
\label{sec:Con}
We have constructed two models based on the DDM mechanism where the $A_4$ is spontaneously broken at the seesaw scale into a remanent $Z_2$. The models have two $Z_2$ odd and three $Z_2$ even RH neutrinos, the latter giving light neutrino masses via type-I seesaw.  The breaking of $A_4$ is leaded by the flavon fields in a way that we get two-zero textures for the light Majorana neutrinos mass matrices. These are in agreement with the experimental data and both mass hierarchies. The models also contain DM candidate the two $Z_2$ odd RH neutrinos. Finally,  we have presented  correlations between mixing parameters and lower bounds for the neutrinoless double beta regions.
\section*{ACKNOWLEDGEMENTS}
I would like to thank my advisor and co-author of the work E. Peinado and the organisers of the 5th Young Researchers Workshop: ''Physics Challenges in the LHC Era" and the XVIII FRASCATI SPRING SCHOOL "BRUNO TOUSCHEK" in Nuclear Subnuclear and Astroparticle Physics for the very interesting events and kind hospitality. This work was possible thanks to the grant No. PAPIIT IA101516 and CONACyT.

\end{document}